\documentclass{reducedws} 

\begin{document}

\title{FREEZE-OUT IN\\ 
ULTRA-RELATIVISTIC HEAVY-ION COLLISIONS\thanks{%
presentation based on work done in collaboration 
with \uppercase{U}rs \uppercase{W}iedemann}}

\author{BORIS TOM\'A\v SIK}

\address{CERN, Theory Division, CH-1211 Geneva 23, Switzerland}

\maketitle

\vspace*{-0.1ex}

\abstracts{
I discuss the effects and quantities that influence the decoupling 
of particles from the fireball. The crucial role is played 
by the scattering rate. I show the results for the scattering
rate at SPS and RHIC and discuss their implications.}

Strongly interacting matter is produced in ultra-relativistic 
collisions of heavy atomic nuclei at SPS and RHIC. 
The system cools down quickly, however, and hadronic spectra 
are formed as soon as when interactions disappear. Hence, they
carry no direct signals of the early hot, dense and 
deconfined phase. There are indirect signatures
of collective behaviour of  strongly interacting matter
encoded in spectra, but it is important to disentangle them from features
generated in the freeze-out process. Therefore,
it is important to understand the freeze-out mechanism 
if one is interested in the early hot plasma phase.


\section{Motivation} 
\label{motiv}

\paragraph{Modelling the freeze-out.}
A particle will decouple from the system when the density 
drops so low that it does not scatter anymore. One expects
that all particles will decouple at roughly the same time,
when the density drops below a critical value.

How big is the space-time region where 
particles are emitted? Cascade generators typically
predict that hadrons are liberated from very early times on, 
during a time interval of up to 50 or 75~fm/$c$. In spite of
that, the standard freeze-out prescription\cite{Cooper:1974mv} 
in hydrodynamic
simulations assumes all particles to be set free at a specified
three-dimensional hyper-surface. I will discuss how good such an
idealization is and under which conditions it works.

In a hydrodynamic model one has to specify the freeze-out
condition. A universal freeze-out criterion, valid for
systems of all sizes at all energies is desirable. A constant particle
freeze-out density has been observed in S-induced reactions at 200~$A$GeV 
projectile energy\cite{na35}. Its universality was disproved by  
measurements at different collision 
energies\cite{Adamova:2002ff}. The pion phase-space density,
also proposed to be universal in the past\cite{Ferenc:1999ku}, shows a 
dramatic increase at RHIC\cite{star}. A criterion using the
critical mean free path\cite{Adamova:2002ff} does not take into 
account the expansion of the fireball; I will discuss this below. 

I approach the problem of a universal freeze-out criterion differently:
I will not ask for a condition to be fulfilled by the system,
but rather for circumstances under which a particle escapes.
The freeze-out condition turns out 
to depend on the particle momentum. This is close to the
cascade-generator understanding of the freeze-out.


\paragraph{Understanding spectra.}
Hadronic spectra are formed at the freeze-out but they still  
contain information about the earlier dynamics of the system.
The collective transverse flow that develops throughout the 
whole history of a collision flattens the spectrum 
and may lead to its concave curvature. 

On the other hand, a freeze-out mechanism in which high-$p_\perp$
particles are liberated before the low-$p_\perp$ ones can also 
make the spectrum concave, since high-$p_\perp$'s would 
come from a hotter region.

This shows that a good understanding of the freeze-out is necessary
for the measurement of the collective flow from the spectra.


\section{Freeze-out depends on \dots}
\label{fout}

\paragraph{\dots\ chemical composition.}
When calculating the scattering rate of a test particle,
the density is multiplied by the cross-section for scattering on particles
from the medium. Densities of different species are always multiplied
by the corresponding  cross-sections. Thus one concludes
that freeze-out is characterized by a critical mean free path (here for pions)
\begin{equation}
\label{mft}
1/\lambda_\pi = 
n_\pi \sigma_{\pi\pi} + n_N \sigma_{\pi N} + n_K \sigma_{\pi K} + \dots
\, .
\end{equation}
An estimate of the pion mean free path\cite{Adamova:2002ff,Tomasik:2002qt} 
at freeze-out gives a value of the order of 1~fm, 
maybe\cite{Tomasik:2002qt} 2--3~fm.

\paragraph{\dots\ expansion.}
If freeze-out was characterized merely by density and chemical composition,
the mean free path would have to be comparable with the system size, which
is about an order of magnitude larger than the observed $\lambda_\pi$.
The expansion of the fireball is crucial\cite{garpman}, however. 
After the pion travels the distance $\lambda_\pi$, the density drops and 
the system
is too dilute for another scattering. The mean free
path is rather short at the moment of the last scattering, but it is 
infinite a couple of fermis later.

This mechanism sets a dynamical condition for freeze-out\cite{garpman}:
it occurs when the rate of density decrease is faster than the 
scattering rate.

\paragraph{\dots\ particle momentum.}
The mean free path and the density decrease
rate are quantities that characterize the medium. On the other 
hand, the dependence on momentum\cite{Grassi:1994ng} 
is specific for every particle 
and leads to the construction of a freeze-out criterion that will treat
individual particles. 

Therefore, I can formulate the freeze-out criterion by saying that 
a particle decouples if its probability
to escape without further scattering is reasonably 
large\cite{Hung:1997du,Sinyukov:2002if} (say, 0.5).

The sharp freeze-out along a three-dimensional hyper-surface 
is a special case, in which the escape probability of {\em all} 
particles quickly changes from  
0 to 1. If, however, the escape probability grows slowly
or has a pronounced momentum dependence, the sharp freeze-out
approximation becomes questionable.


\section{Formalism, calculation, and results}
\label{form}

The probability of a particle with momentum $p$  at the space-time point 
$(\tau,x)$
not to scatter anymore is given as\cite{Tomasik:2002qt,Grassi:1994ng}
\begin{equation}
\label{esprob}
P(x,\tau,p) = \exp\left( - \int_\tau^\infty d\tau' \,
R(x+v\tau',p) \right ) \, .
\end{equation}
The (opacity) integral of the scattering rate $R(x+v\tau',p)$ along the 
expected trajectory of the particle gives the average number
of collisions the particle would suffer if it moved straight.

The density and the chemical composition enter into the calculation
of the scattering rate. The expansion dynamics determines its 
time dependence through the time depenednce  of the density.

I show results\cite{Tomasik:2002qt} for the scattering rate 
at the lower bound of the opacity integral and compare values for SPS and
RHIC. Data on the pion freeze-out phase-space density at RHIC\cite{star}
show an increase by factor of 2 with respect to the SPS, so it is interesting 
to compare the two systems and ask
how it is possible that pions cease to interact at RHIC in a system denser 
than that  at the SPS?

In calculating\cite{Tomasik:2002qt} 
the results, a thermal distribution of momenta was
assumed and the abundances of individual species were tuned 
in order to reproduce the observed phase-space densities and 
ratios of mid-rapidity yields.

\begin{figure}[t]
\centerline{\epsfxsize=3.9in\epsfbox{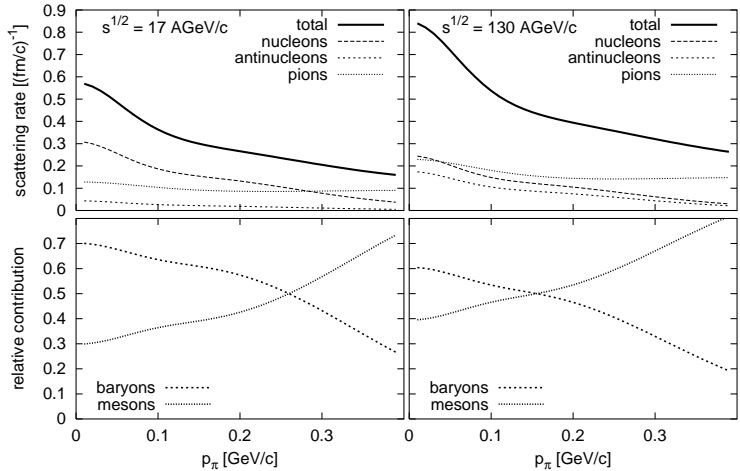}}   
\caption{Scattering rate as a function of pion momentum 
with respect to the medium with $T=100$~MeV. Results are obtained
for SPS (left) and RHIC (right). Contributions to the scattering 
rate from scattering on pions, nucleons and antinucleons 
are indicated. The two lower panels shows the baryonic and mesonic relative
contributions.
\label{fig1}}
\end{figure}
Typically, the nucleon contribution to pion scattering is smaller at RHIC 
as there is less baryon stopping, but it is roughly replaced by scattering
on antinucleons (Fig.~\ref{fig1}). 
Scattering on pions is stronger at RHIC than at the SPS,
because of the increase in phase-space density, but in neither case
does it clearly dominate the scattering rate because of the  small 
$\pi\pi$ cross-section. The dramatic effect of high phase-space density 
thus has little influence on the freeze-out.

\begin{figure}[t]
\centerline{\begin{minipage}[b]{2.48in}
\epsfxsize=2.18in\epsfbox{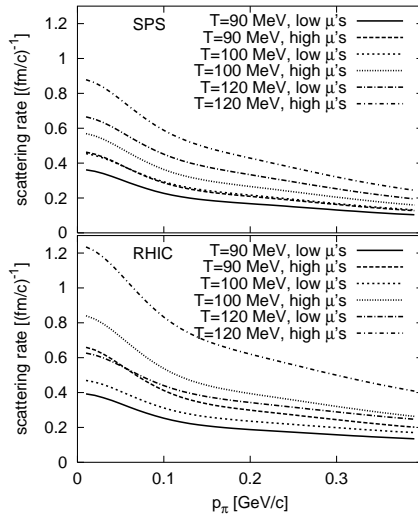}
\end{minipage}
\begin{minipage}[b]{1.6in}   
\caption{The pion scattering rates as functions of momentum
with respect to the medium, calculated for different temperatures
and sets of chemical potentials allowed by data. Details can 
be found in the original paper\protect\cite{Tomasik:2002qt}.
\newline
\label{fig2}}
\end{minipage}}
\end{figure}
All results are summarized in Figure~\ref{fig2}. 
It is seen that generally the scattering rate strongly depends
on the momentum. High-$p_\perp$ particles decouple more easily from the 
system and may be able to escape earlier when the system is
still rather dense. Hence, there is {\em time ordering} in particle 
production and it does not seem to be a good approximation to
assume that {\em all} particles are produced at a single freeze-out
hyper-surface (as done in many hydrodynamic simulations). 

The scattering rate also increases with the temperature, even if the
density is kept constant. The escape probabilities for realistic 
density decrease rates were estimated\cite{Tomasik:2002qt} from 
(\ref{esprob}); it was found that for a temperature of 120~MeV
the chance to escape is about 10\% for a particle with momentum above 
250\, MeV. In order to obtain a reasonable escape probability,
say 30--50\%, the temperature must drop to about 100~MeV.
This suggests a freeze-out at a low temperature.

\section{Conclusions}
\label{conc}

Time ordering of the emission of different momenta may have observable
effects on the $K_\perp$ dependence of HBT radii, because high-$p_\perp$ 
particles would come from a smaller source than the low-$p_\perp$ 
ones. The bulk of the pions seem to freeze-out rather late, 
at a temperature of about $100$~MeV.


\end{document}